\documentclass[conference]{IEEEtran}
\IEEEoverridecommandlockouts
\usepackage{amsmath,amsfonts,amssymb}
\usepackage{algorithmic}
\usepackage{algorithm}
\usepackage{array}
\usepackage[caption=false,font=normalsize,labelfont=sf,textfont=sf]{subfig}
\usepackage{textcomp}
\usepackage{stfloats}
\usepackage{url}
\usepackage{verbatim}
\usepackage{graphicx}
\usepackage{cite}
\usepackage{orcidlink}
\usepackage{booktabs} 
\usepackage{multirow} 
\usepackage{float}    
\begin{document}

\title{Enhanced Web Payload Classification Using WAMM: An AI-Based Framework for Dataset Refinement and Model Evaluation}

\author{
\begin{tabular}{ccc}
\textbf{Heba Osama}~\orcidlink{0000-0003-4550-9356} &
\textbf{Omar Elebiary} &
\textbf{Youssef Qassim} \\[2pt]
\small heba.osama@cyshield.com &
\small omar.elebiary@cyshield.com &
\small youssef.qassim@cyshield.com \\[2pt]
\small Cyshield Company, Cairo, Egypt &
\small Cyshield Company, Cairo, Egypt &
\small Cyshield Company, Cairo, Egypt \\[10pt]

\textbf{Mohamed Amgad} &
\textbf{Ahmed Maghawry}~\orcidlink{0000-0002-9559-2939} &
\textbf{Ahmed Saafan} \\[2pt]
\small mohamed.amgad@cyshield.com &
\small ahmed.maghawry@cyshield.com &
\small ahmed.saafan@cyshield.com \\[2pt]
\small Cyshield Company, Cairo, Egypt &
\small Cyshield Company, Cairo, Egypt &
\small Cyshield Company, Cairo, Egypt \\[10pt]

\multicolumn{3}{c}{
\begin{tabular}{c}
\textbf{Haitham Ghalwash}~\orcidlink{0000-0001-8887-2716} \\[2pt]
\small haitham.ghalwash@coventry.ac.uk \\[2pt]
\small Coventry University – Egypt Branch 
\end{tabular}
}
\end{tabular}
}

\IEEEpubid{%
\makebox[0pt][r]{%
\raisebox{-1.6\baselineskip}{
\parbox[b]{\columnwidth}{\footnotesize
This work has been submitted to the IEEE for possible publication.\\
Copyright may be transferred without notice, after which this version\\
may no longer be accessible.
}}}}

\maketitle
\IEEEpubidadjcol

\begin{abstract}
Web applications increasingly face evasive and polymorphic attack payloads, yet traditional web application firewalls (WAFs) based on static rule sets such as the OWASP Core Rule Set (CRS) often miss obfuscated or zero-day patterns without extensive manual tuning. This work introduces WAMM, an AI-driven multiclass web attack detection framework designed to reveal the limitations of rule-based systems by reclassifying HTTP requests into OWASP-aligned categories for a specific technology stack. WAMM applies a multi-phase enhancement pipeline to the SR-BH 2020 dataset that includes large-scale deduplication, LLM-guided relabeling, realistic attack data augmentation, and LLM-based filtering, producing three refined datasets. Four machine and deep learning models are evaluated using a unified feature space built from statistical and text-based representations. Results show that using an augmented and LLM-filtered dataset on the same technology stack, XGBoost reaches 99.59\% accuracy with microsecond-level inference while deep learning models degrade under noisy augmentation. When tested against OWASP CRS using an unseen augmented dataset, WAMM achieves true positive block rates between 96 and 100\% with improvements of up to 86\%. These findings expose gaps in widely deployed rule-based defenses and demonstrate that curated training pipelines combined with efficient machine learning models enable a more resilient, real-time approach to web attack detection suitable for production WAF environments.
\end{abstract}

\begin{IEEEkeywords}
Web application security, HTTP payload classification, Multiclass detection, Machine learning, XGBoost, Web attacks, Web application firewall.
\end{IEEEkeywords}

\section{Introduction}\label{sec1}
\IEEEPARstart{W}{eb} applications represent essential components of recent services, covering sectors from finance and e-commerce to healthcare and cloud-based platforms. These applications, however, have become primary targets for increasingly sophisticated attackers who create malicious HTTP requests to exploit vulnerabilities, including SQL injection (SQLi), cross-site scripting (XSS), and local file inclusion (LFI). Although traditional web application firewalls (WAFs), such as those utilizing the Open Web Application Security Project (OWASP) Core Rule Set (CRS), offer an initial layer of protection through signature-based rules, they are limited in their effectiveness against obfuscated payloads or zero-day vulnerabilities, often resulting in misclassification and reliance on outdated rule sets \cite{ref1}. Furthermore, effective deployment demands significant manual adjustments to optimize the balance between security and usability, which increases operational complexity and overhead.

WAFs continue to function as a crucial component of web security, even though recent incidents have exposed systemic vulnerabilities within their defenses. Empirical studies indicate that inconsistencies in request parsing \cite{ref2} and the use of corrupted or mutated payloads \cite{ref3} can bypass even commonly deployed platforms, including Amazon Web Services (AWS), Azure, Cloudflare, and ModSecurity, with evasion rates reaching up to 79\% \cite{ref3}. These findings highlight that signature-based defenses are insufficient against evolving threats.

The integration of artificial intelligence (AI) within WAFs has demonstrated significant transformative impacts, and academic research increasingly shows the importance of integrating machine learning (ML) and deep learning (DL) into WAFs. Adaptive rule generation \cite{ref4}, reinforcement learning-based analysis \cite{ref5}, and large language model (LLM)-driven payload synthesis \cite{ref6} have shown advances in detection accuracy, reduced false positives, and increased resilience to zero-day exploits \cite{ref7}. Such advances indicate the emergence of a new generation of intelligent WAFs capable of evolving static signature-based detection.

Despite these advances, numerous significant challenges remain. Initially, accurately representing HTTP payloads for classification purposes continues to be an unresolved challenge. Second, numerous studies concentrate on binary detection (malicious versus benign) rather than on comprehensive multiclass classification aligned with the OWASP Top 10 categories. Finally, the challenge of integrating ML models into production WAF systems without incurring excessive latency or resource consumption remains unresolved.

To address these gaps, WAMM is produced as an AI-driven framework designed to classify HTTP requests into various attack types aligned with OWASP standards, trained on a specific technology stack. The WAMM framework highlights four principal strengths:
\begin{itemize}
    \item Enhancement of the high-quality dataset through the refinement of the SR-BH dataset via cleansing, relabeling using LLMs, and realistic augmentation of both malicious and benign traffic sources.
    \item Explicitly categorize and map payloads to OWASP-aligned attack types which include SQLi, XSS, LFI, remote file inclusion (RFI), and command injection (CMDi).
    \item Comprehensive benchmarking analysis for various models was performed such as extreme gradient boosting (XGBoost), bidirectional long short-term memory (BiLSTM), convolutional neural network and bidirectional long short-term memory (CNN-BiLSTM), and distilled bidirectional encoder representations from transformers (DistilBERT). These models are selected for their respective strengths to address the challenges of sparse, sequential, and semantic data representation.
    \item Direct benchmarking the effectiveness of WAMM against OWASP CRS to establish industry baselines.
\end{itemize}

To conduct a systematic evaluation of the performance and design choices behind WAMM, this paper is driven by the following key research questions:
\begin{itemize}
    \item RQ1: To what extent does WAMM effectively cover and identify the web attack categories delineated within the OWASP Top 10?
    \item RQ2: Which featurization approach for web attack payloads is appropriate for machine learning models?
    \item RQ3: Which ML and DL models demonstrate the highest effectiveness in multiclass classification of web attacks?
    \item RQ4: How does the performance of WAMM generalize to both known and unseen attack payloads, including zero-day patterns?
    \item RQ5: Which composition of training data supports generalizable web attack detection more effectively?
    \item RQ6: How is the accuracy of WAMM evaluation when compared to the OWASP Core Rule Set for a specific technology stack?
\end{itemize}

The rest of the paper is organized as follows. Section two reviews related research on WAFs and the enhancement of web attack detection using ML. Section three provides the WAMM methodology, including dataset enhancement, preprocessing, feature representation techniques, and training strategy. The experimental setup, dataset versions, evaluation metrics, and deployment configuration are detailed in section four. Section five presents the experimental evaluation and results discussion of the classification performance. Finally, Section Six concludes the presented study and highlights potential future directions for the ongoing development of the WAF system.

\section{Related Work}\label{sec2}

\subsection{Web-Based Attack Detection: Overview and Challenges}
Web attacks are becoming harder to spot because many malicious payloads now arrive obfuscated inside legitimate HTTP requests. These payloads frequently include SQL injection attempts \cite{ref8, ref9}, cross-site scripting \cite{ref10}, and command injection, all of which continue to target sectors such as healthcare, government, and critical industrial systems. Recent deep learning approaches have shown strong performance in detecting these threats. For example, E-WebGuard combines CNN and BiLSTM layers that reported accuracy above 99\% on the SR-BH 2020 dataset \cite{ref11}, and another CNN-LSTM model designed for SQLi and XSS achieved comparable precision \cite{ref12}. Other work includes a CNN-only classifier trained on CSIC2010v2 \cite{ref13}, a deception-oriented system that analyzes cookies in real time and reached 99.94\% accuracy \cite{ref14}, and a three-stage GAN-Transformer workflow that reported 99.97\% on HTTP URLs \cite{ref15}.

These approaches each come with their own tradeoffs. Convolutional recurrent hybrids capture both fine-grained local patterns and broader temporal behavior \cite{ref11, ref12}, whereas CNN-only models prioritize simplicity \cite{ref13}. Systems that are based on deception push detection further by engaging with adversarial behavior in real time \cite{ref14}, and multi-stage frameworks aim for stepwise optimization across several components \cite{ref15}. Despite these advances, several limitations remain. Many studies still rely on legacy or synthetic datasets such as CSIC2010, which reduces the realism of their evaluations. Latency and resource consumption are often left unreported; models focused solely on SQLi or XSS provide limited multiclass coverage, and small or outdated corpora reduce resilience against adversarial manipulation. WAMM is designed to address these gaps. It uses lightweight classifiers such as XGBoost alongside a heavily augmented dataset that includes encoding variants, obfuscation techniques, and a broader set of attack vectors. The framework is tailored to a specific technology stack, which helps it balance detailed detection performance, stack-specific behavior, and practical deployment efficiency.

\subsection{Traditional Detection Techniques}
Conventional intrusion detection systems and WAFs typically rely on static signatures and handcrafted rules, as seen in tools like Snort, Suricata, and ModSecurity. These traditional systems usually require extensive tuning before deployment to keep false positives at acceptable levels. These systems perform well against known threats but struggle with zero-day attacks. Furthermore, from a practical perspective, most organizations lack the specialized resources required for continuous fine-tuning. As a result, they often rely on a one-time tuning service which may be accepted for the protected and ready-made application but becomes challenging for in-house systems that undergo frequent updates and therefore require ongoing adjustments to prevent false positives and ensure a smooth user experience. This study \cite{ref1} proposed ModSec-Learn, augmenting ModSecurity with supervised learning and sparse regularization to refine rules and improve precision. Another \cite{ref2} introduced WAFFLED, revealing 1,207 bypasses across major platforms due to HTTP parsing inconsistencies and proposing HTTP-Normalizer to address them. A review \cite{ref16} highlighted limitations in datasets, inconsistent labeling, and underuse of URL semantics. Another study \cite{ref17} designed a two-phase NIDS with feature selection and ensemble ML, reaching 99.8\% accuracy on CICIDS2017 with real-time validation.

These contributions show complementary directions: ModSec-Learn enhances rule-based detection but remains tied to CRS \cite{ref1}, WAFFLED exposes structural flaws overlooked by adaptive models \cite{ref2}, anomaly detection reviews identify systemic issues while offering guidance and a best practices framework without a turnkey methodology for intrusion detection \cite{ref16}, and NIDS research focuses on deployment-level evaluation but at the flow rather than payload level \cite{ref17}. Key gaps remain in relying on single datasets, coarse binary outputs, and the lack of real-world performance metrics. WAMM addresses these by training on a combined, balanced multi-dataset corpus tailored to a specific technology stack to improve generalization, providing fine-grained multi-class classification for more actionable insights.

\subsection{Machine Learning Approaches to Web Attack Detection}
To overcome the limitations of static rules, researchers have adopted ML and DL for adaptive, data-driven detection. CNN–RNN hybrids have shown strong performance in multiclass payload classification \cite{ref11, ref12}, while attention-based fusion of URL and payload features improves contextual understanding \cite{ref18}. Distributed ML-based WAF architectures reduce processing time by up to 18 times without losing accuracy \cite{ref19}. Furthermore, using rule-based genetic algorithms decreases latency \cite{ref20}. Graph-based XSS modeling with GCNs enhances structural representation \cite{ref21}, whereas incorporating knowledge-graph features into classical learners yields notable gains in accuracy and area under the curve \cite{ref22}. Flow-level reconnaissance detection effectively identifies scanners such as ZAP and Nikto \cite{ref23}, and for HTTPS traffic, a SuperLearner ensemble achieves over 97\% fingerprinting accuracy while its obfuscation defender reduces attack success to 2.89\% with minimal overhead \cite{ref33}.

It was also observed that many studies relied on outdated datasets such as CSIC2010 and its derivatives, leading to potential performance overestimation. Other studies targeted specific attack families while relying on curated side information or omitting payload inspection and latency evaluation, which limits real-world applicability \cite{ref21, ref22, ref23}. WAMM extends these efforts by curating time-aware datasets while integrating hybrid URL payload features and reporting standardized robustness and latency metrics that balance semantic fidelity with operational performance.

\subsection{Dataset Limitations in Prior Work}
Moreover, previous studies showed that the efficiency of ML and DL for web attack detection is sensitive to dataset quality, diversity, realism; however, many benchmarks have structural and semantic flaws that limit generalization and stumble deployment. Surveys have documented multi-label inconsistencies and showed that multi-label methods outperform single-label ones \cite{ref24}. Work on SQLi used LSTMs to address class imbalance \cite{ref8}, while other studies reported strong results on legacy corpora such as CSIC 2010 \cite{ref11, ref12}. Additional efforts built CNN- and RNN-based payload classifiers on narrow or ISP-sourced data \cite{ref25, ref26}. A multi-scale CNN-BiLSTM with attention achieved near-perfect XSS metrics \cite{ref27}, and dataset analyses highlighted unrealistic traffic generation and labeling noise \cite{ref28}. SQLi-specific TextCNN-BiLSTM with BERT features improved recognition but remained single-vector \cite{ref29}, while temporal and TF-IDF features boosted XSS accuracy without adversarial evaluation \cite{ref30}. Finally, a benchmarking framework showed that common HTTP datasets omit families such as PHP web shells, limiting real-world coverage \cite{ref31}.

Across this literature, recurring issues persist as severe class imbalance skews metrics and calibration, limited payload diversity weakens resilience to encoding and obfuscation, label noise and redundancy undermine ground truth, synthetic traffic produces unrealistic distributions that miss multi-vector campaigns, and narrow coverage excludes critical classes such as web shells \cite{ref31}. These issues and limitations will negatively affect reported performance and raise doubts about operational reliability. Recent worldwide events such as the MOVEit and Zimbra incidents, showed how obfuscated polymorphic payloads can bypass WAFs, leading to a gap between benchmark results and production realities. To merge this gap, WAMM effectively extends SR-BH 2020 with both augmentation and curation in the form of a pipeline that applies structural deduplication to cut redundancy in addition to Burp Suite-driven injections to add realistic noisy traffic in addition to transformation operators to generate encoding and obfuscation variants aligned with observed gaps, followed by LLM-assisted validation to preserve label semantics. The resulting datasets showed improved label fidelity and attack coverage robustness to polymorphic threats, hence bridging the evaluation deployment gap.

\subsection{Real-Time and Low-Latency Detection Models}
In addition to accuracy, IDSs must meet strict real-time requirements in latency-sensitive domains such as e-government, healthcare, and high-throughput e-commerce platforms. Previous research investigated architectures that combine high fidelity with low latency. GAN-based augmentation improved robustness on CIC-IDS2017 by synthesizing attack traffic \cite{ref34}. PhishingRTDS, a containerized BiLSTM-attention framework, achieved 99\% F1 with real-time deployment \cite{ref35}. APELID, a parallel ensemble IDS with adversarial refinement, achieved remarkably high F1 on CSE-CIC-IDS2018 and NSL-KDD \cite{ref36}. At the web layer, cookie semantics were integrated into detection pipelines \cite{ref37}. Infrastructure-oriented work explored fog-cloud partitioning to reduce delays \cite{ref38}, line-rate detection with programmable P4 switches \cite{ref39}, and lightweight spiking-CNN hybrids for high-throughput inference \cite{ref40}. Collectively, these studies show the feasibility of scaling IDS toward real-time operation.

Limitations remain in prior work: GAN-based frameworks and ensembles mainly target network-layer traffic rather than HTTP payload semantics; PhishingRTDS and similar systems are narrowly scoped without multiclass OWASP coverage; fog-cloud and hardware accelerators improve throughput but neglect semantic fidelity; and spiking-CNN models emphasize efficiency at the cost of deep payload analysis. Consequently, existing methods only partially balance semantic accuracy with latency constraints.

To address this, WAMM integrates payload-level semantics with real-time classification by extending SR-BH 2020 through deduplication, adversarial payload generation, and LLM validation. Latency-efficient inference modules were prepared for deployment by exporting the model as a serialized object using the pickle (PKL) format, and this resource was consumed by a FastAPI service for extended testing as well as production integration, enabling multiclass OWASP-aligned detection with microsecond-scale performance suitable for production-grade WAF deployment.

\section{Proposed Model}\label{sec3}

\subsection{Overview}
The WAMM is a comprehensive and extensible framework designed to detect and classify HTTP payloads into common attack pattern enumeration and classification (CAPEC) which aligns attack categories, with a focus on the most popular categories based on OWASP Top 10, such as SQLi, XSS, and CMDi, with focused training on a given technology stack. It integrates ML and DL techniques with a multi-phase data enhancement pipeline applied to the SR-BH 2020 dataset \cite{ref41}. This includes cleansing, relabeling via LLMs, and payload augmentation using tools that generate penetration testing payloads. The system is developed to support scalable deployment in real-time environments and offers standardized benchmarking across models and dataset versions.

\subsection{System Workflow}
\begin{figure*}[!t]
\centering
\includegraphics[width=\textwidth]{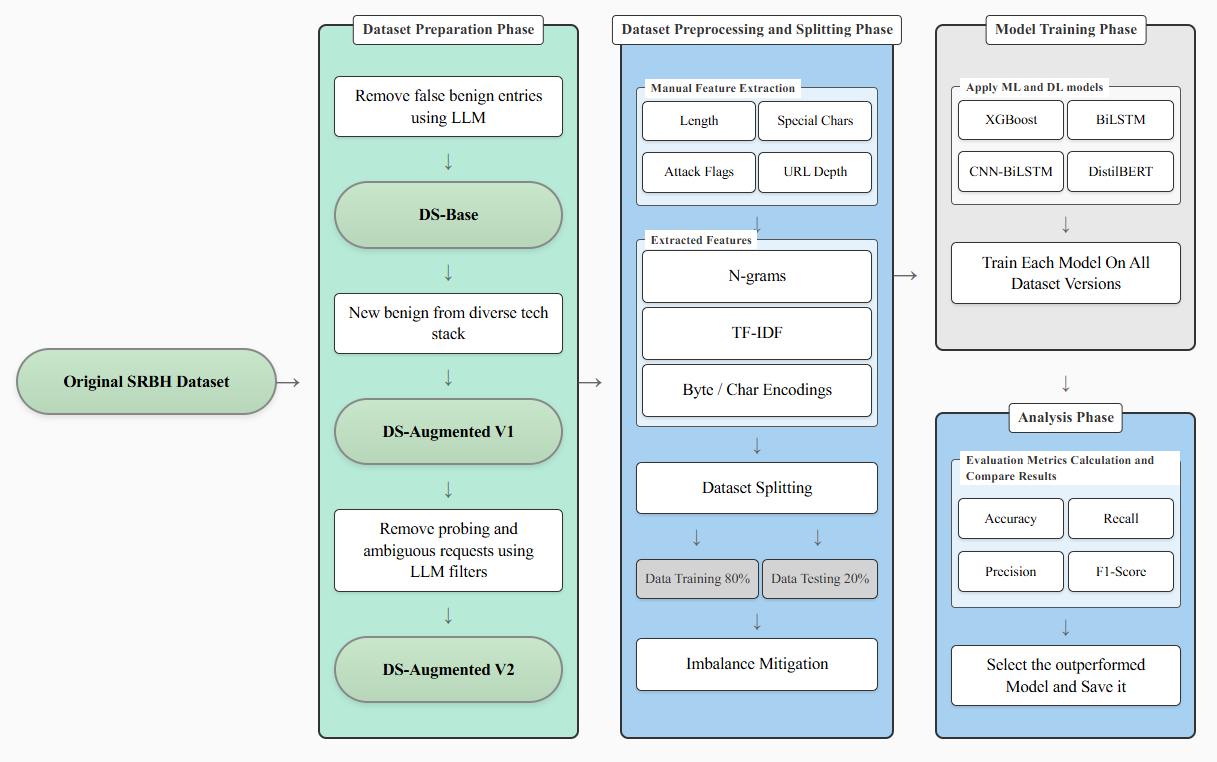}
\caption{WAMM Framework Pipeline}
\label{fig1}
\end{figure*}

Our framework follows a four-phase pipeline: dataset preparation, preprocessing and splitting, model training and evaluation, and comparative analysis, which is designed for scalability and robust performance across diverse web attack categories. Figure \ref{fig1} describes the overall processing pipeline, showing dependencies between the four phases.

Each stage plays a critical role in ensuring high-quality model performance and generalizability across OWASP attack categories.
Phase 1 models the process of dataset preparation where comprehensive data cleansing, taxonomy modernization and systematic quality enhancement are done. This phase addresses fundamental dataset quality issues through systematic fixes including deduplication supported by LLM-guided mislabel correction, in addition to controlled request augmentation strategies. Phase 2 models the process of preprocessing and splitting where standardized feature engineering was implemented in addition to dataset partitioning and class balancing techniques to ensure consistent input representations across all model architectures while preserving the distributional characteristics of real-world traffic patterns.
Phase 3 models the process of model training that applies four different training approaches: XGBoost, Bidirectional LSTM, CNN-BiLSTM, and DistilBERT under identical experimental conditions. To comprehensively evaluate systematic performance comparison.
And finally, phase 4 models the process that provides systematic evaluation using relevant metrics, testing and deployment feasibility analysis based on inference speed for each model to ensure practical applicability in production environments where near-realtime processing may be required.

\subsection{Dataset Preparation}
As a result of our framework described above, the WAMM dataset has been structured into three distinct versions:
\begin{itemize}
    \item \textbf{DS-Base}: This is the foundational dataset that was created through a thorough improvement process of the original SR-BH 2020’s data issues.
    \item \textbf{DS-Augmented-v1}: The base dataset was enhanced with the augmentation of a large sample of new malicious payloads and a greater variety of benign network traffic.
    \item \textbf{DS-Augmented-v2}: For this version an LLM was used to filter DS-Augmented-v1 resulting in only correctly classified malicious samples being retained.
\end{itemize}
These improved datasets were created to address critical quality issues identified in the original SR-BH 2020 data. This was done by three steps of the validation process in which automated pattern recognition, LLM-guided classification, and final manual verification by experts were combined.

\subsubsection{Automated pattern detection}
A comprehensive fingerprinting script was developed with 127 regular expression patterns across six attack categories to identify potential mislabeled samples in SR-BH 2020. The patterns include classic attack signatures like ('$\backslash$s*(union$|$UNION)$\backslash$s+(select$|$SELECT) for SQLi), many obfuscation techniques custom to each attack and possible encoding variations (\%2e\%2e\%2f for path traversal). This automated screening revealed approximately 10\% mislabeling within the benign class only, providing quantitative evidence of dataset quality issues that were subsequently validated through LLM analysis. Table \ref{tab1} presents the per attack type breakdown of mislabeled records flagged by the script.

\begin{table}[!t]
\caption{Script’s Result: SR-BH 2020 Mislabeled Benign Data}\label{tab1}
\centering
\footnotesize 
\setlength\tabcolsep{3pt} 
\begin{tabular}{@{} l p{3cm} @{}} 
\toprule
Attack Type & Mislabeled Rows Count \\
& (\% of Benign Mislabeled Requests) \\
\midrule
CMDi & 16,740 (3.348\%) \\
Server-Side Template Injection (SSTI) & 2,500 (0.500\%) \\
SQLi & 20,800 (4.160\%) \\
XSS & 2,335 (0.467\%) \\
Path Traversal & 8,000 (1.600\%) \\
\bottomrule
\end{tabular}
\end{table}

\subsubsection{LLM-guided classification}
\begin{figure}[!t]
\centering
\includegraphics[width=\columnwidth]{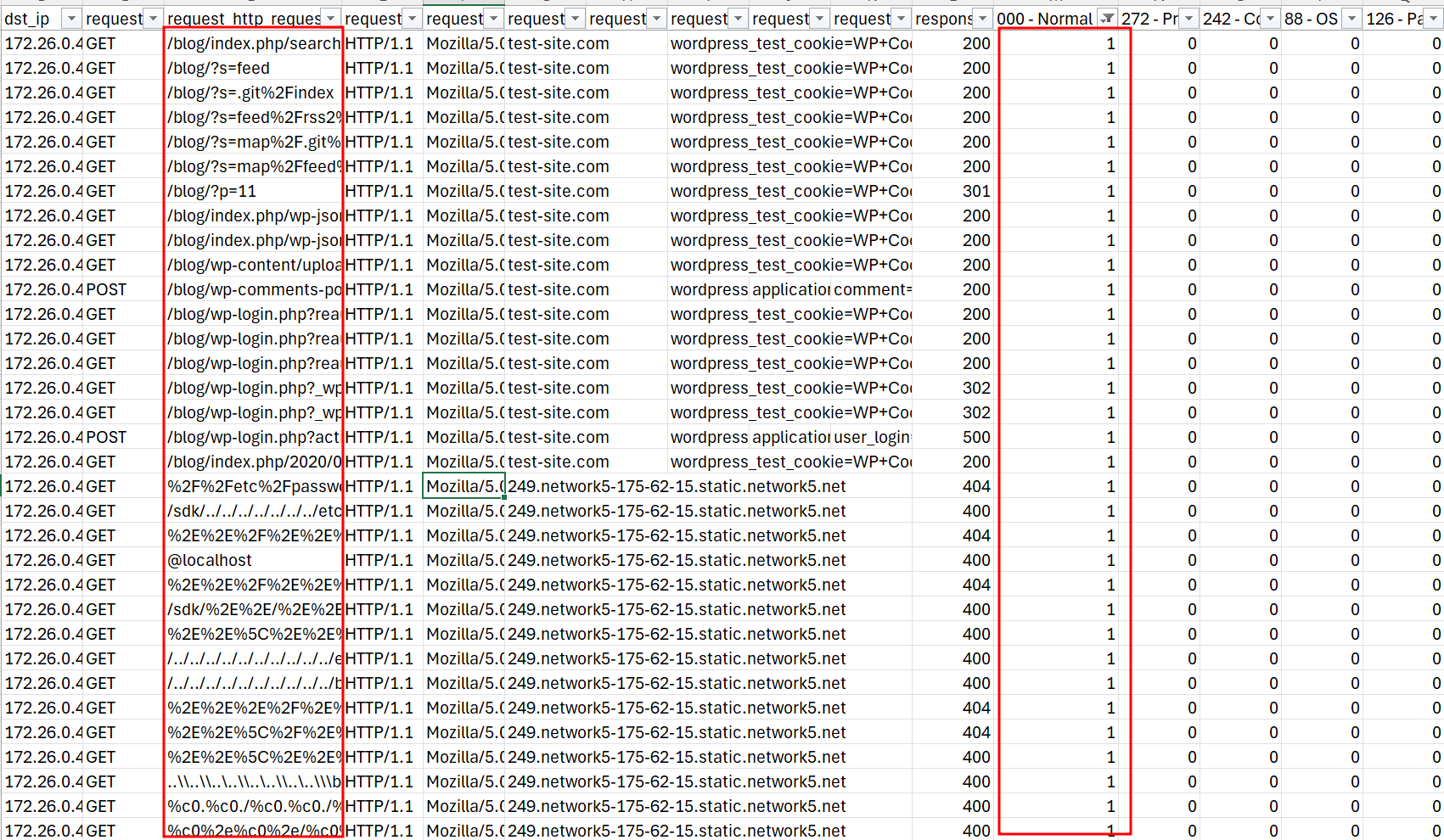}
\caption{SR-BH 2020 Mislabelled Benign Requests}
\label{fig2}
\end{figure}

The full benign class from SR-BH 2020 represented by 525,195 samples was processed using the Qwen3-8B model \cite{ref42} with a tailored security analysis prompt designed to identify the OWASP Top 10 attack patterns. The LLM evaluation identified 48,522 malicious requests incorrectly labeled as benign, confirming the automated analysis results with higher precision. Figure \ref{fig2} presents samples of SR-BH 2020 benign-labeled requests that were flagged as mislabels.

\subsubsection{Expert manual verification}
Over 300 randomly selected samples from different classes after running both automated and LLM classifications were manually reviewed by experts in web penetration testing, achieving 100\% concordance with LLM classifications. The introduced validation step introduced confidence in the LLM-guided approach, while confirming the systematic nature of the mislabeling instances.

A systematic evaluation of all thirteen CAPEC attack categories revealed some major inconsistencies. Several categories failed to meet definitional criteria: HTTP Request Smuggling (CAPEC-33) lacked the required content-length or transfer-encoding headers to carry the attack, with no way to identify the requests from normal traffic, thus making detection impossible; dictionary-based password attacks (CAPEC-16) were indistinguishable from normal login HTTP requests too; and HTTP verb tampering (CAPEC-274) relied on response content rather than request-based indicators. Mislabeling was also identified as the "fake the source of data" (CAPEC-194) class incorrectly contained SSRF attacks rather than any actual fake source of data attacks, and CMDi (CAPEC-248) contained only a single undocumented entry. After removing the troubled categories and correcting mislabeled samples the refined dataset retained five reliable classes: Protocol Manipulation represented by 9,153 samples, OS Command Injection is represented by 7,482 samples, Path Traversal is represented by 20,992 samples, SQLi is represented by 250,311 samples normal traffic with 525,195 samples ensuring greater reliability for training and classification.

\begin{figure*}[!t]
\centering
\includegraphics[width=\textwidth]{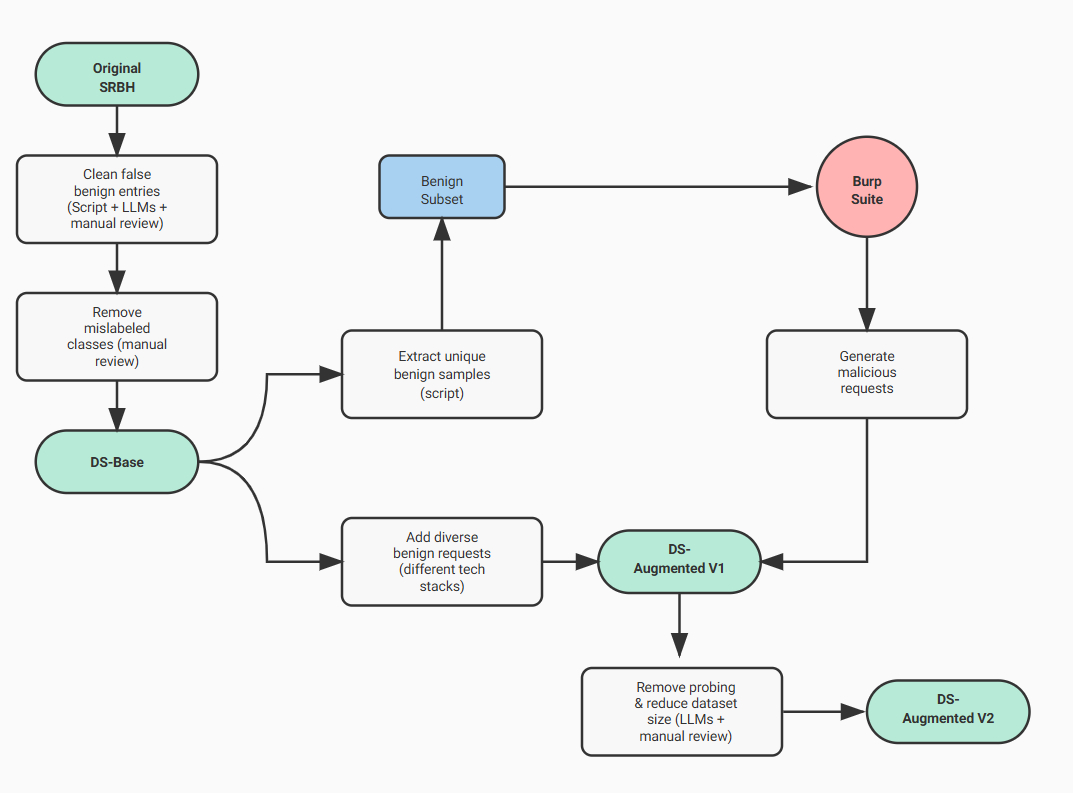}
\caption{Dataset Enhancement Pipeline}
\label{fig3}
\end{figure*}

To address these limitations, the DS-Base was carefully developed utilizing a comprehensive validation pipeline as shown in Figure \ref{fig3}. Furthermore, MinHash and locality-sensitive hashing (LSH) were applied to eliminate any structurally redundant benign entries producing a fairly representative subset suitable for augmentation.

To further enrich the improvement process of the DS-Base, the DS-Augmented-v1 was created to improve attack diversity and enrich traffic realism. For this, synthetic malicious payloads were generated using Burp Suite scans to ensure coverage of the OWASP Top 10 categories. On the other hand, benign traffic was enriched with legitimate requests from platforms such as Facebook, Amazon, and Medium to better capture real-world patterns and diversify the target technology stack of the dataset’s requests.

Furthermore, the DS-Augmented-v2 extends the refinement of this approach through LLM-powered validation to remove probing and noisy payloads. It provides a cleaner and more reliable training set for high-stakes detection by retaining only high-certainty malicious samples from the previous activity. This prevents possible overfitting influenced by any possible tool-generated noise to ensure better generalization that suits production systems.

The refined collection of datasets supports a robust training pipeline while maintaining realistic and actionable traffic patterns \cite{ref43, ref44}. Table \ref{tab2} shows the dataset enhancement pipeline, outlining the key transformation phases applied to the original SR-BH dataset.

\begin{table*}[!t]
\caption{Dataset Enhancement Pipeline Summary.}\label{tab2}
\centering
\resizebox{\textwidth}{!}{%
\begin{tabular}{lp{6cm}p{5cm}}
\toprule
Phase & Description & Tools/Techniques \\
\midrule
Phase 1: Cleansing & LSH was used to remove about 519K redundant benign payloads, and corrected around 48K labels using LLM (Qwen3-8B). & Datasketch: A Python library for probabilistic data structures, used here for LSH to remove duplicates \cite{ref45}. Qwen3-8B: A large language model applied to correct mislabeled data through context-aware classification \cite{ref49} \\
Phase 2: Taxonomy Update & Updated class labels for alignment with OWASP and the 2025 threat landscape (added XSS, SSTI, and SSRF). & Mapping Rules: A custom rule set maps old dataset labels to the updated OWASP-aligned taxonomy \cite{ref47}. \\
Phase 3: Burp Augmentation & Custom scans per attack class, generated malicious traffic. & Burp Suite Pro: Industry-standard tool for web vulnerability scanning and custom attack payload generation \cite{ref48}. \\
Phase 4: Benign Diversity & Simulated benign requests from real-world apps like Amazon, Facebook, and Medium. & Burp Suite Pro + Logger++: Burp extension Logger++ captures and exports HTTPS traffic for dataset enrichment \cite{ref46}. \\
Phase 5: Versioning & DS-Augmented-v1: raw payloads (including probes) - DS-Augmented-v2: LLM-confirmed malicious requests only. & SQLite + local LLM pipeline \cite{ref50}. \\
\bottomrule
\end{tabular}%
}
\end{table*}

\subsection{Preprocessing and Splitting}
The preprocessing and splitting stage transformed raw HTTP request data into a structured input representation suitable for model training comprising three components: feature engineering, dataset partitioning, and class imbalance handling.
For the feature engineering component, a couple of feature categories were derived \cite{ref63}. The first introduced handcrafted attributes such as payload length, counts of special characters including ( $<$, $>$, ", ', ;), binary indicators of attack patterns including (SQL keywords, path traversal), counts of numeric and percent-encoded characters, Shannon entropy, URL depth, and unique character counts. These features are described in detail in Table \ref{tab3}. These patterns combined together succeeded to capture low-level obfuscation and encoding signals. The second category utilized the automated text analysis approach of TF-IDF weighting to highlight rare tokens \cite{ref32, ref51}, N-grams to model local attack patterns including ( "../", "SELECT", "?id=1") and byte- or character-level encodings to preserve raw structure. Combining both manual and automated features succeeded in yielding semantically enriched and structurally robust representations of attack vectors.

\begin{table}[!t]
\caption{Features Description}\label{tab3}
\centering
\begin{tabular}{@{}lp{3.5cm}@{}}
\toprule
Feature & Purpose \\
\midrule
Payload length & Detects anomalously long inputs that may indicate injection attacks \\
Character ratios & Identifies suspicious symbol frequencies (quotes, angle brackets, special characters) \\
Shannon entropy & Measures randomness, indicative of encoded or obfuscated attack payloads \\
URL depth & Flags deeply nested path structures that may indicate path traversal attempts \\
Unique character counts & Captures diversity of characters used, helping identify unusual patterns or obfuscation techniques \\
\bottomrule
\end{tabular}
\end{table}

To achieve fair and balanced evaluation, each dataset was systematically partitioned into 80\% of the data used for training while the remaining 20\% of the data used for testing subsets. For this, stratified sampling was applied to preserve the attack categories distribution , while maintaining normal traffic across subsets to ensure representative covering.
On the other hand, certain attack types occur significantly less frequently than others in the SR-BH dataset, for this, such resulting class imbalance was addressed by calculating class weights inversely proportional to the frequency of each class within the training set \cite{ref52}. And finally, these weights were integrated into the learning process to reduce bias toward majority classes while enhancing detection performance for underrepresented attack categories.

\subsection{Classification Models}
The WAMM pipeline contains four distinct classification approaches to capture complementary aspects of malicious web payload detection. Based on the literature review and our experience, XGBoost was selected as our baseline ensemble method, leveraging TF-IDF vectorization combined with automatically engineered features such as payload length and manually engineered features like custom patterns for each attack type \cite{ref53}. This tree-based methodology offers computational efficiency and interpretability, effectively addressing the sparse, high-dimensional feature spaces characteristic of textual security data.

Furthermore, a BiLSTM network was implemented to capture sequential patterns and positional dependencies by processing raw byte-level payload representations \cite{ref54}. Such bidirectional architecture enables exploitation of both forward and backward contextual information introducing an effective mechanism for detecting the sensitivity of attack patterns.

Additionally the hybrid CNN-BiLSTM architecture integrates the strengths of convolutional and recurrent approaches where the convolutional layers extract local n-gram patterns and spatial features \cite{ref55}. In addition, the BiLSTM layers capture long-range dependencies and create a comprehensive framework that identifies both localized and distributed attack signatures.

Finally, DistilBERT was adapted using a knowledge-distilled transformer model through fine-tuning for binary payload classification \cite{ref56}. This approach introduces a mechanism to process complete HTTP payloads using contextualized token embeddings to achieve superior semantic understanding of malicious intent while maintaining the needed computational efficiency for real-time WAF deployment.

\subsection{Model Training and Evaluation}
In the introduced WAMM approach, by reaching this point, the dataset is considered to be preprocessed and appropriately balanced; hence, the next stage focuses on training a range of candidate supervised ML and DL models followed by systematic evaluation.
Each candidate model is trained adopting the same standardized input features and class-weighted loss functions to guarantee fair and balanced comparison. The training process is executed on controlled CPU configurations to guarantee reproducibility and deployment viability in resource-limited settings. Furthermore, early stopping and learning rate scheduling techniques are utilized where possible to handle overfitting and improve convergence.
Each candidate model performance \cite{ref57} is evaluated through a combination of classification and efficiency metrics that are aligned with the main objectives of this research.
Accuracy reflects the overall proportion of correctly identified requests out of the total dataset as shown in equation (\ref{eq:acc}):
\begin{equation}
\label{eq:acc}
Accuracy = \frac{TP + TN}{TP + TN + FP + FN}
\end{equation}
Where TP (True Positives) represents malicious requests that are correctly identified by the model, TN (True Negatives) denotes benign requests that are correctly recognized by the model, FP (False Positives) refers to the benign requests misclassified by the model as attacks, and FN (False Negatives) indicates malicious requests that are incorrectly labeled as normal by the model.

The recall metric represents the true positive instances rate, it captures the fraction of actual attacks successfully detected, the recall metric calculation method is shown in equation (\ref{eq:recall}):
\begin{equation}
\label{eq:recall}
Recall = \frac{TP}{TP+FN}
\end{equation}
The precision metric measures the reliability of attack predictions by computing the fraction of correctly identified attacks among all targeted requests labeled as malicious, the precision metric calculation method is shown in equation (\ref{eq:prec}):
\begin{equation}
\label{eq:prec}
Precision = \frac{TP}{TP+FP}
\end{equation}
On the other hand, F1-Score provides the harmonic mean of precision and recall, balancing the trade-off between them as shown in equation (\ref{eq:f1}):
\begin{equation}
\label{eq:f1}
F1\text{-}Score = \frac{2 * Precision * Recall}{Precision + Recall}
\end{equation}
Model performance was assessed using classification and computational efficiency metrics aligned with the research objectives. Precision, recall, and F1-score (both per each class and macro averaged) were used to measure detection effectiveness across OWASP Top 10 categories, satisfying the first research question, confusion matrices were also generated to visualize each model’s behavior. Inference time was recorded as the average latency per HTTP request showing real world usage speed and addressing the fourth research question concerning real time feasibility.
All candidate models were evaluated against the same uniform stratified split to ensure fair and balanced comparison. WAMM was also benchmarked against the OWASP CRS to address the sixth research question. This methodology enabled systematic comparison of ML and DL models to address the third research question by highlighting differences in architectural performance and providing further insights regarding detection robustness, deployment suitability and external validity.

\section{Experimental Setup}\label{sec4}
All experiments (except for DistilBERT models) were conducted on a CPU-only server with these specifications: 8 vCPUs, 32 GB RAM, 250 GB SSD, and Ubuntu 22.04 LTS.
DistilBERT was trained separately on a GPU-equipped machine with a NVIDIA RTX 4060, and 8 GB VRAM to account for the higher computational demands of Transformer models. This setup provides realistic evaluation for both lightweight and computationally demanding models \cite{ref58}.

We built our training system in Python using PyTorch and Hugging Face Transformers for DL models, and XGBoost with Scikit-learn for classical ML. Data handling relied on Pandas, NumPy, and SciPy, with visualization performed using Matplotlib and Seaborn. Custom modules supported preprocessing, training, and evaluation workflows.

Four dataset variants were used: DS-Original (SR-BH 2020), DS-Base, DS-Augmented-v1, and DS-Augmented-v2. Each included Full\_request (HTTP query/payload) and class (attack label, SQLi, XSS, RFI, Normal).

Preprocessing involved text normalization, conversion of payloads to strings, and removal of unlabeled samples. Feature engineering combined statistical attributes as shown in table 3 with semantic indicators. Regex patterns for XSS, SQLi, RFI, LFI, SSTI, RCE, CMDi, and path traversal captured domain-specific signatures \cite{ref59}.

Four models were trained: XGBoost, BiLSTM, CNN-BiLSTM, and DistilBERT. As shown in Table \ref{tab4}, XGBoost used TF-IDF with engineered features, and BiLSTM and CNN-BiLSTM processed character-level inputs (ASCII-encoded and padded to 100), with CNN-BiLSTM adding a 1D convolutional layer. DistilBERT was fine-tuned on tokenized request strings (max length 128) using the official tokenizer \cite{ref60}.

\begin{table*}[!t]
\caption{ML and DL Models Pipeline Summary.}\label{tab4}
\centering
\resizebox{\textwidth}{!}{%
\begin{tabular}{llll}
\toprule
Model & Input Format & Tokenization / Features & Description \\
\midrule
XGBoost & TF-IDF + Manual & Char n-grams (1,2), max 2000 & Classical ML with sparse input representation \\
BiLSTM & Character-level string & ASCII-encoded padded sequences & Sequential RNN-based model \\
CNN-BiLSTM & Character-level string & Convolution + BiLSTM combination & Captures both local and long-range patterns \\
DistilBERT & Raw text & Tokenized using the Hugging Face model & Transformer-based model for contextual understanding \\
\bottomrule
\end{tabular}%
}
\end{table*}

To ensure comparability, all models were trained and evaluated using the same stratified train-test split.
Class imbalance was handled by estimating class weights with Scikit-learn’s compute\_class\_weight utility and incorporating those weights during training for classical algorithms, and by using class-weighted loss functions for our neural networks \cite{ref61}. For a fair and reproducible comparison between deep models, shared hyperparameters were kept consistent where possible, and only adjusted architecture-specific settings to reflect each model’s requirements. For example, the BiLSTM and the CNN-BiLSTM experiments used the same character-level input length, embedding dimension, and bidirectional recurrent layers, while DistilBERT was trained using the standard tokenization and fine-tuning workflow for pretrained transformers. The principal training settings for every model are summarized in Table \ref{tab5}.

\begin{table}[!t]
\caption{Training Hyperparameters for Deep Learning Models}\label{tab5}
\centering
\begin{tabular}{@{} l p{2cm} p{2cm} @{}}
\toprule
Parameter & BiLSTM / CNN-BiLSTM & DistilBERT \\
\midrule
Epochs & 5 & 3 \\
Batch Size & 64 & 16 \\
Learning Rate & Adam (default) & 2e-5 (AdamW) \\
Max Sequence Length & 100 & 128 \\
Embedding Dimension & 100 & Pretrained \\
Hidden Layers & 2 (bidirectional) & 1 (pretrained) \\
\bottomrule
\end{tabular}
\end{table}

Furthermore, once all candidate models are executed, the highest-performing model will be targeted for an extended evaluation against the OWASP CRS for accuracy to further benchmark the proposed model against one of the field’s standard benchmarking methodologies.

\section{Experimental Evaluation and Results Discussion}\label{sec5}
This section evaluates the WAMM web attack detection framework across four datasets and models. Regarding detection coverage, feature representation, model performance, generalization, dataset composition, and comparison with the OWASP CRS, this analytical study addresses six research questions.
The evaluation progresses from analyzing dataset characteristics to visualizing model behavior, concluding with performance comparisons and architectural trade-offs.

\subsection{Dataset Characteristics and Visualization}
\begin{figure*}[!t]
\centering
\includegraphics[width=\textwidth]{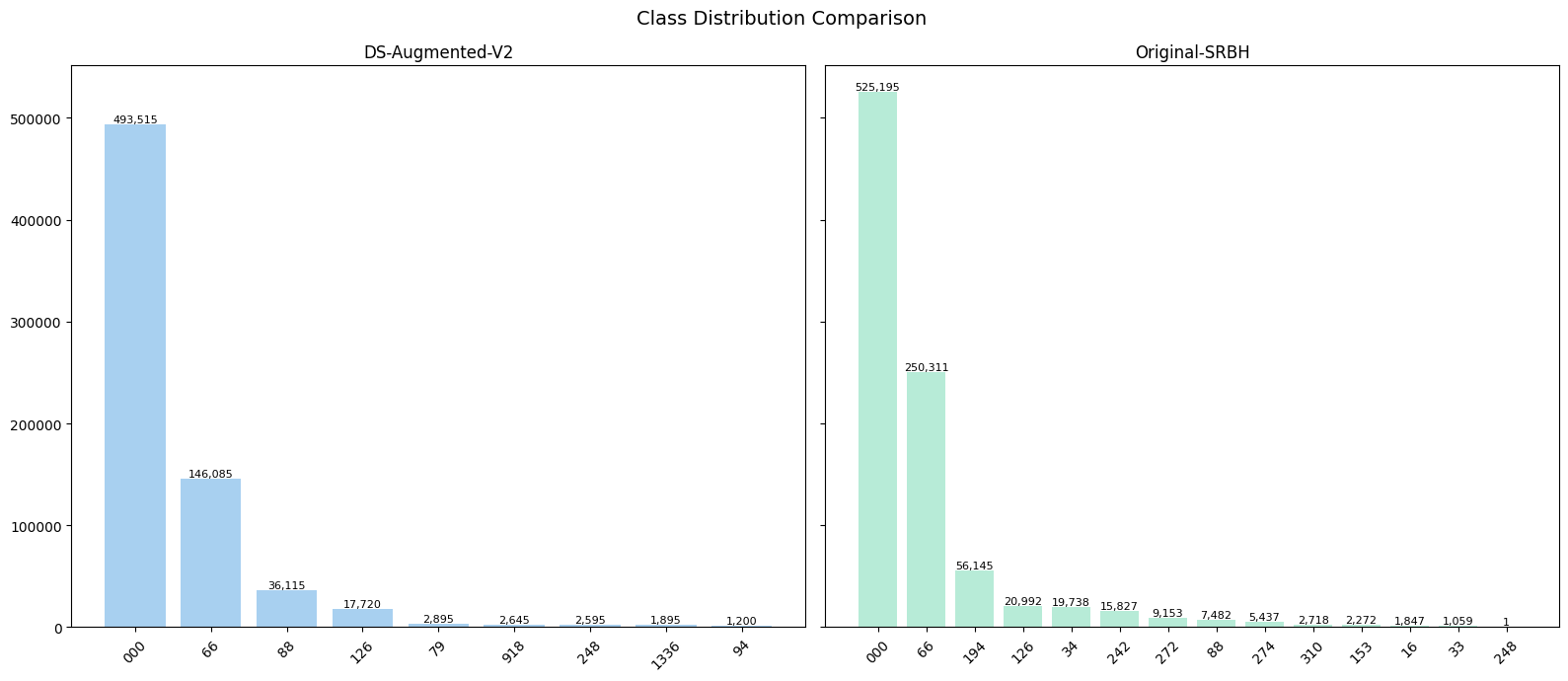}
\caption{DS-Augmented-v2 and DS-Original Label Distribution}
\label{fig4}
\end{figure*}

The distribution of class labels and feature characteristics was analyzed across four datasets: DS-Original (SR-BH 2020), DS-Base, DS-Augmented-v1, and DS-Augmented-v2. This analysis was performed to evaluate how impressively the datasets covered OWASP categories and to understand their overall composition. This analysis relates to the first research question concerning OWASP coverage and the fifth research question regarding how dataset design affects performance. Figure \ref{fig4} shows the class distributions for both DS-Original and DS-Augmented-v2.

The datasets demonstrate a robust alignment with OWASP categories (SQLi, XSS, LFI, RFI, CMDi, SSTI, etc.), which ensures that WAMM covers a comprehensive range of realistic payloads, especially DS-Augmented-v2. Augmented samples generated by the Burp Suite and relabeled with the assistance of LLM improve coverage across various attack categories. Although class imbalance remains (SSTI), the usage of class weighting reduces issues related to the minority class, thereby addressing the first research question.

DS-Augmented-v1 adds diversity but includes noisy tool-generated probes that hinder training. This is addressed in DS-Augmented-v2 through LLM-guided filtering, which improves feature coherence, class balance, and overall model performance, highlighting the value of semantically validated synthetic data for multiclass classification and answering the fifth research question.

\begin{figure}[!t]
\centering
\includegraphics[width=\columnwidth]{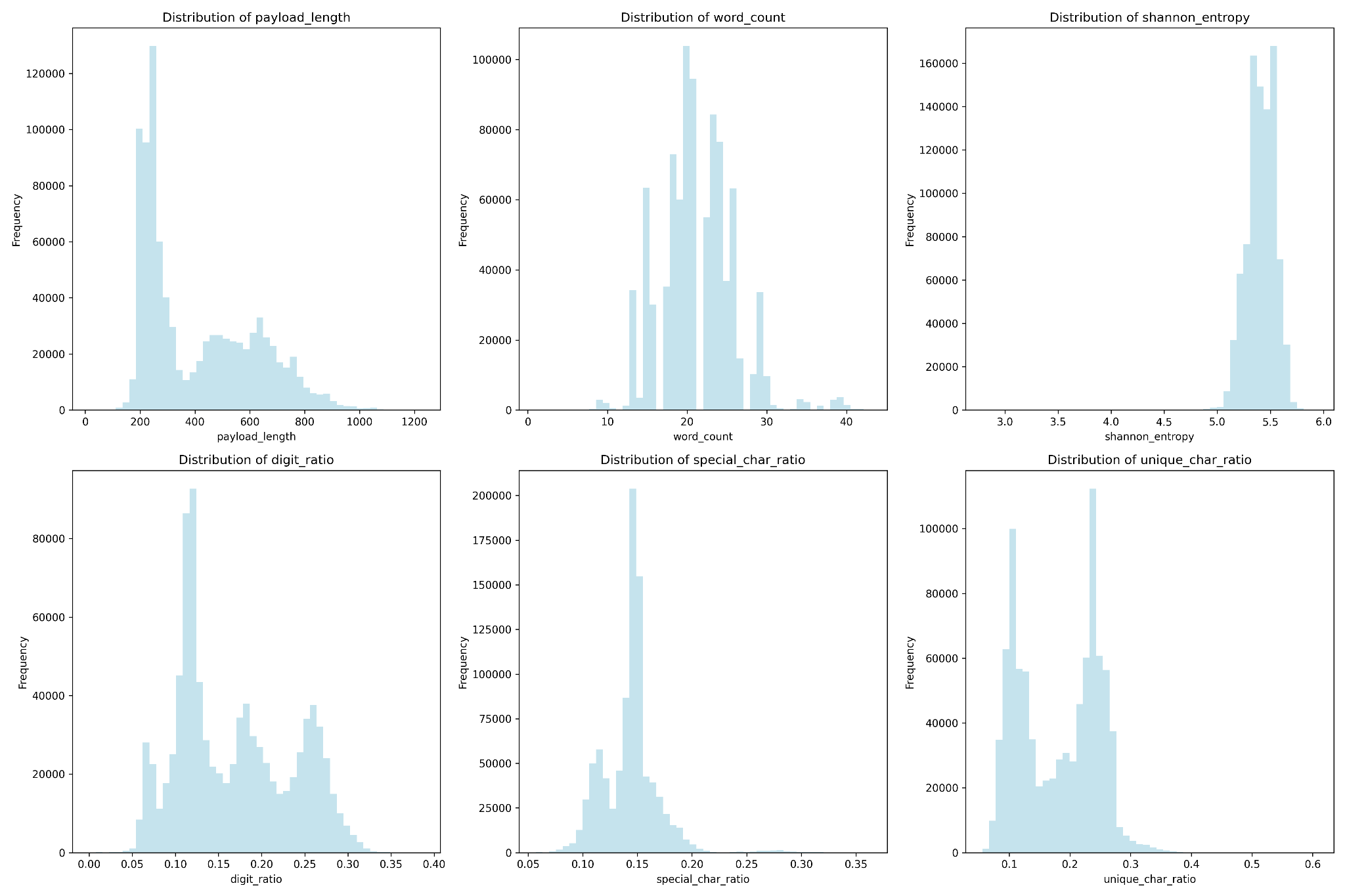}
\caption{DS-Original Feature Distribution}
\label{fig5}
\end{figure}

\begin{figure}[!t]
\centering
\includegraphics[width=\columnwidth]{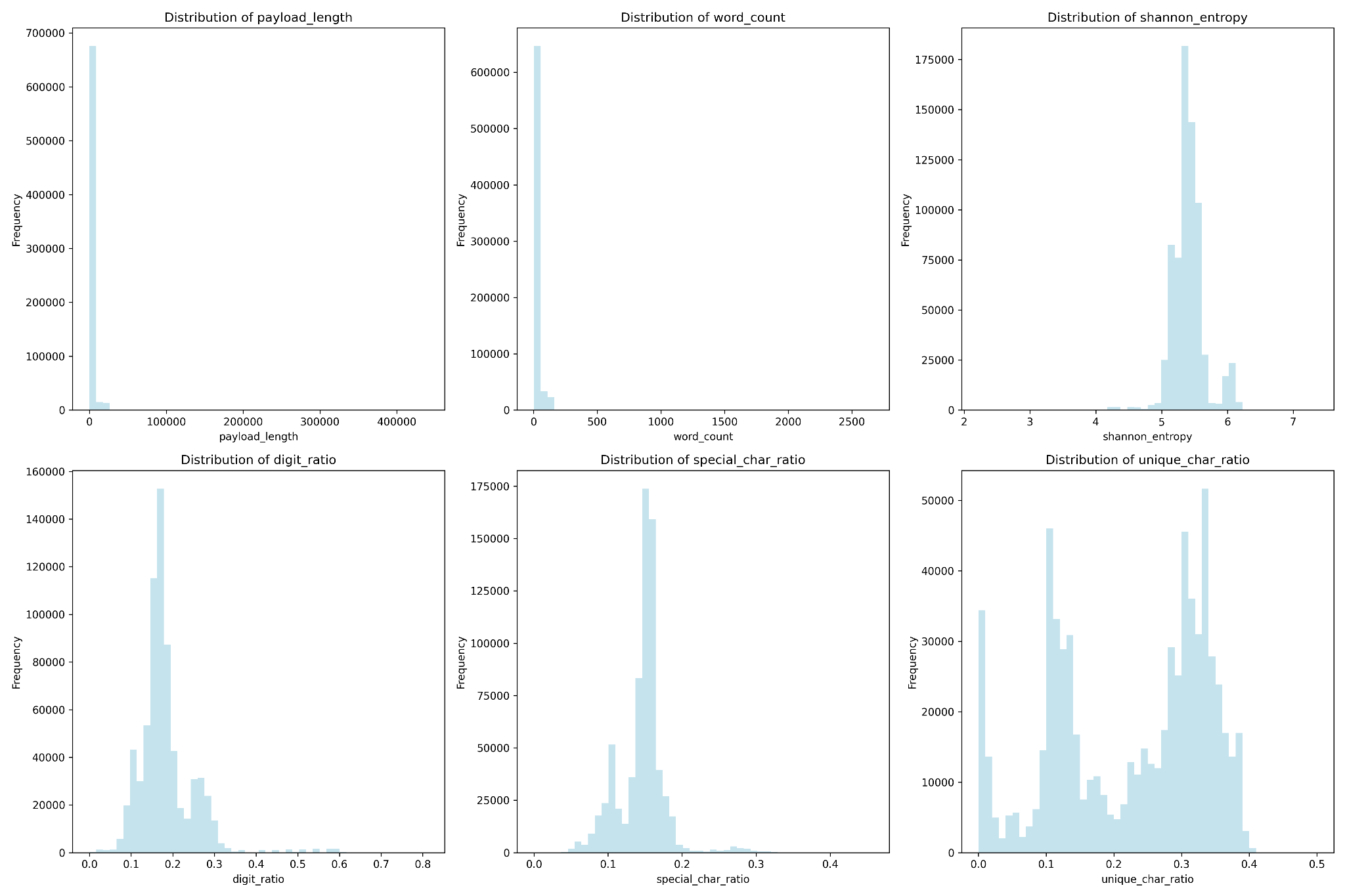}
\caption{DS-Augmented-v2 Feature Distribution}
\label{fig6}
\end{figure}

Statistical analysis of payloads (payload\_length, word\_count, shannon\_entropy, digit\_ratio, special\_char\_ratio) is shown in Figures \ref{fig5} and \ref{fig6}. DS-Augmented-v2 shows enhanced structural consistency, while DS-Original and DS-Base reflect heavy benign real-world traffic. DS-Augmented-v1 shows increased variability due to synthetic probes, whereas DS-Augmented-v2 produces more consistent profiles following filtering of low-confidence and malformed samples.

\subsection{Model Performance Visualization}
\begin{figure}[!t]
\centering
\includegraphics[width=\columnwidth]{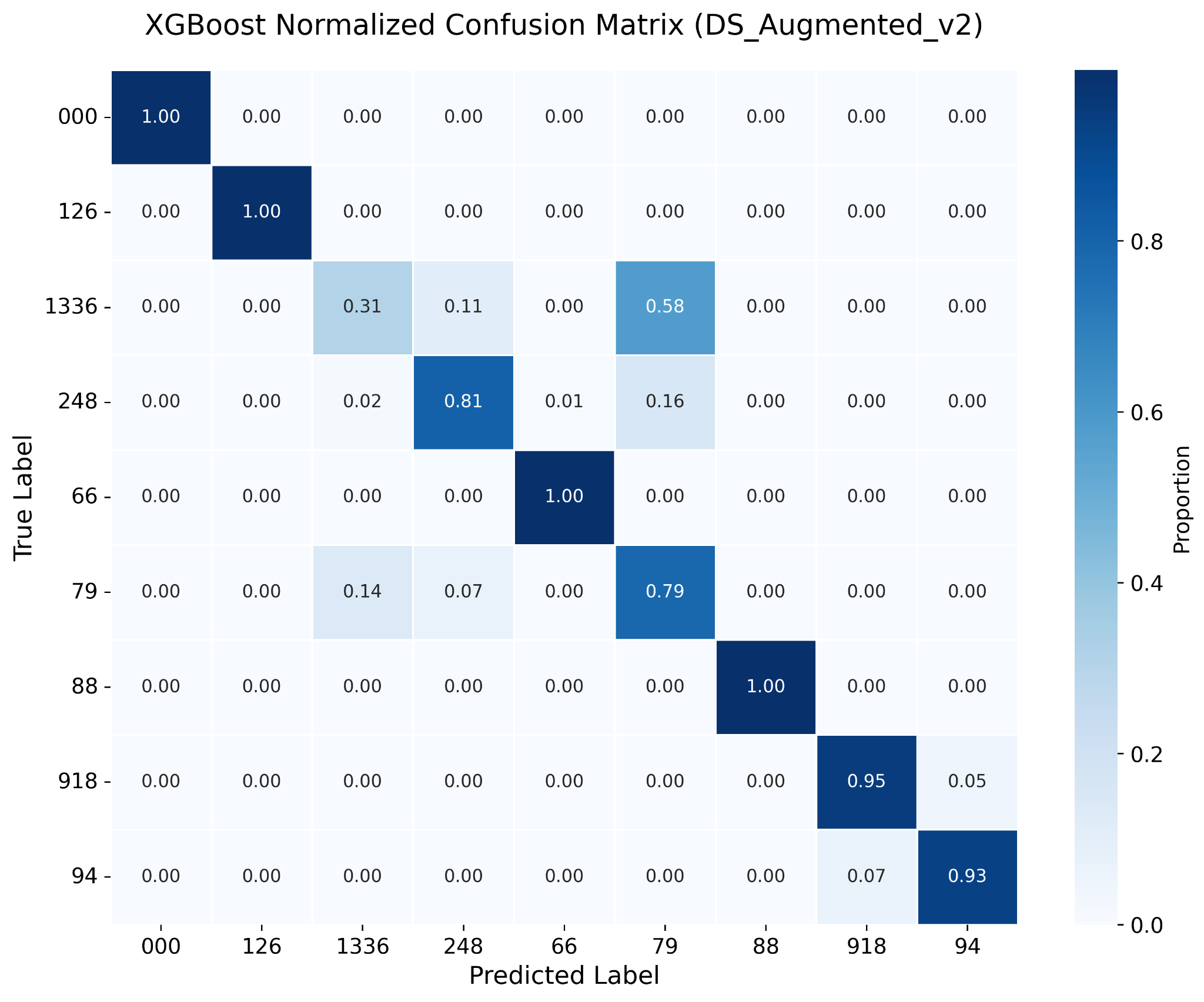}
\caption{Confusion Matrix of XGBoost for Ds\_Augmented-v2}
\label{fig7}
\end{figure}

Confusion matrices were generated for all models on each dataset to address the first research question concerning class-wise detection and the fourth research question regarding generalization to unseen patterns. As an example, Figure \ref{fig7} presents the XGBoost confusion matrix for DS-Augmented-v2.

XGBoost consistently delivers high precision and recall on the most common OWASP classes such as SQLi and CMDi, demonstrating robustness against the most frequent attacks. DistilBERT and BiLSTM show weaker performance on minority or syntactically complex classes like SSTI and XSS, especially in DS-Augmented-v1, highlighting sensitivity to class imbalance and complexity, addressing our first research question.

The declined performance observed on DS-Augmented-v1, caused by its large volume of probe-heavy synthetic payloads, reflects the challenge of generalizing to noisy or obfuscated variants and directly relates to the fourth research question. By comparison, DS-Augmented-v2 restores performance across all models, showing that curated augmentation improves generalization. This confirms WAMM’s ability to exploit high-quality synthetic data to simulate zero-day and mutated attacks, underscoring the importance of dataset quality for multiclass web payload detection.

\subsection{Model Overall Performance}
Table \ref{tab6} represents the summary of a detailed classification report for each model and dataset combination. XGBoost consistently had the best accuracy across all datasets, with a high of 99.59\% on DS-Augmented-v2 and a low of 99\% on DS-Original and DS-Base. BiLSTM and CNN-BiLSTM performed similarly, achieving around 97.2\% accuracy on DS-Original and DS-Base, while their accuracy declined to around 87\% on DS-Augmented-v1. DistilBERT demonstrated strong performance on DS-Original, achieving 98.45\% accuracy, but dropped to 85.45\% on DS-Augmented-v1, indicating sensitivity to distributional shifts and augmentation noise.

\begin{table*}[!t]
\caption{Summary of Model Classification Performance Across All Datasets}\label{tab6}
\centering
\resizebox{\textwidth}{!}{%
\begin{tabular}{llllll}
\toprule
Dataset & Model & Accuracy & F1-Score & Train Time & Inference Time \\
& & & & (HH:MM) & (ms) \\
\midrule
SRBH-Original & XGBoost & 99.25\% & 96.77\% & 0:15 & 0.071 \\
 & BiLSTM & 97.23\% & 77.11\% & 4:10 & 0.713 \\
 & CNN-BiLSTM & 97.12\% & 75.22\% & 5:26 & 0.741 \\
 & DistilBERT & 98.45\% & 85.47\% & 39:19:00 & 14.423 \\
\midrule
DS-Base & XGBoost & 99.01\% & 97.05\% & 0:11 & 0.061 \\
 & BiLSTM & 97.21\% & 81.73\% & 4:01 & 0.711 \\
 & CNN-BiLSTM & 97.11\% & 81.58\% & 4:19 & 0.727 \\
 & DistilBERT & 96.75\% & 75.88\% & 4:29 & 14.500 \\
\midrule
DS-Augmented-v1 & XGBoost & 95.09\% & 77.61\% & 0:25 & 0.051 \\
 & BiLSTM & 87.11\% & 50.28\% & 5:03 & 0.713 \\
 & CNN-BiLSTM & 87.03\% & 51.49\% & 5:45 & 0.738 \\
 & DistilBERT & 85.45\% & 51.30\% & 5:34 & 15.136 \\
\midrule
DS-Augmented-v2 & XGBoost & 99.59\% & 86.07\% & 0:11 & 0.049 \\
 & BiLSTM & 97.56\% & 59.85\% & 3:11 & 0.712 \\
 & CNN-BiLSTM & 97.42\% & 57.61\% & 3:31 & 0.748 \\
 & DistilBERT & 96.64\% & 59.48\% & 3:40 & 14.701 \\
\bottomrule
\end{tabular}%
}
\end{table*}

The macro-averaged F1-score supplies a balanced view by giving equal weight to each class. As shown in Table \ref{tab6}, XGBoost outperformed all other models, achieving F1-scores above 96\% on DS-Original and DS-Base and 86\% on DS-Augmented-v2, while deep models failed to exceed 60\%. DS-Augmented-v1 produced the weakest outcomes through BiLSTM and CNN-BiLSTM results less than 52\% and DistilBERT at 51.3\%, which demonstrated the effect of noisy augmentation. These outcomes highlight the robustness of XGBoost across datasets and justify the second research question.

XGBoost’s strength lies in integrating TF-IDF with engineered features such as payload length, keyword counts, and Shannon entropy, which produces a compact yet effective representation. In contrast, the BiLSTM, CNN-BiLSTM, and DistilBERT rely on byte-level or embedding-based inputs that capture semantics but are noise-sensitive and computationally heavier. Across datasets, XGBoost consistently achieved the highest level of accuracy and efficiency, supporting the third research question, while deep models performed well on clean data but degraded under noisy augmentation. The performance drop on DS-Augmented-v1 underscores the difficulty of generalizing to probe-based payloads relating to the fourth research question, whereas DS-Augmented-v2 demonstrates the value of filtered, semantically validated augmentation for resilience against zero-day patterns.

\begin{figure}[!t]
\centering
\includegraphics[width=\columnwidth]{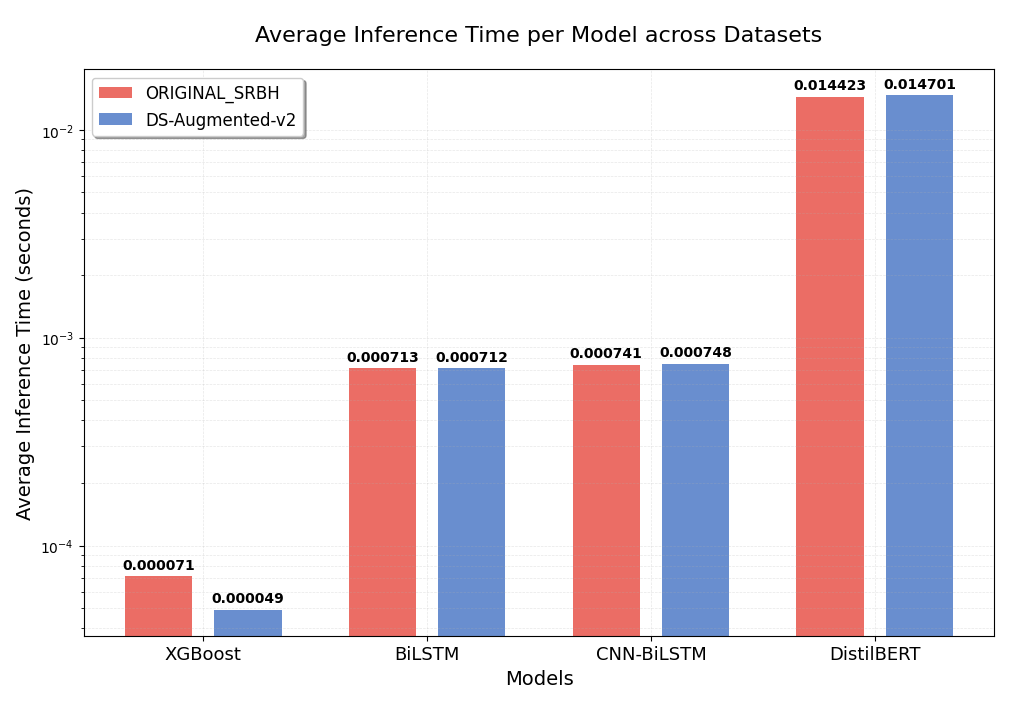}
\caption{DS-Augmented-v2 Inference Speed for All Models}
\label{fig8}
\end{figure}

Training and inference times further highlight trade-offs. XGBoost was trained in less than 30 minutes on all datasets, compared to the BiLSTM and CNN-BiLSTM that required approximately 3 to 5.5 hours and over 39 hours for DistilBERT on DS-Original. Inference followed the same pattern as shown in Figure \ref{fig8}, with XGBoost requiring 49 to 71 microseconds per sample, BiLSTM and CNN-BiLSTM taking around 713 to 748 microseconds, and DistilBERT 14.4 milliseconds, over 140 times slower than XGBoost. These results emphasize the efficiency of XGBoost for low-latency detection versus the higher computational cost but richer context modeling of deep learning approaches.

XGBoost’s short training time and fast inference make it perfect for real-time deployment in WAFs or security monitoring. These results address our third research question, showing that effective model selection must balance accuracy with computational efficiency in operational cybersecurity use cases.

\subsection{Extended WAMM vs. CRS Evaluation}
\begin{figure}[!t]
\centering
\includegraphics[width=\columnwidth]{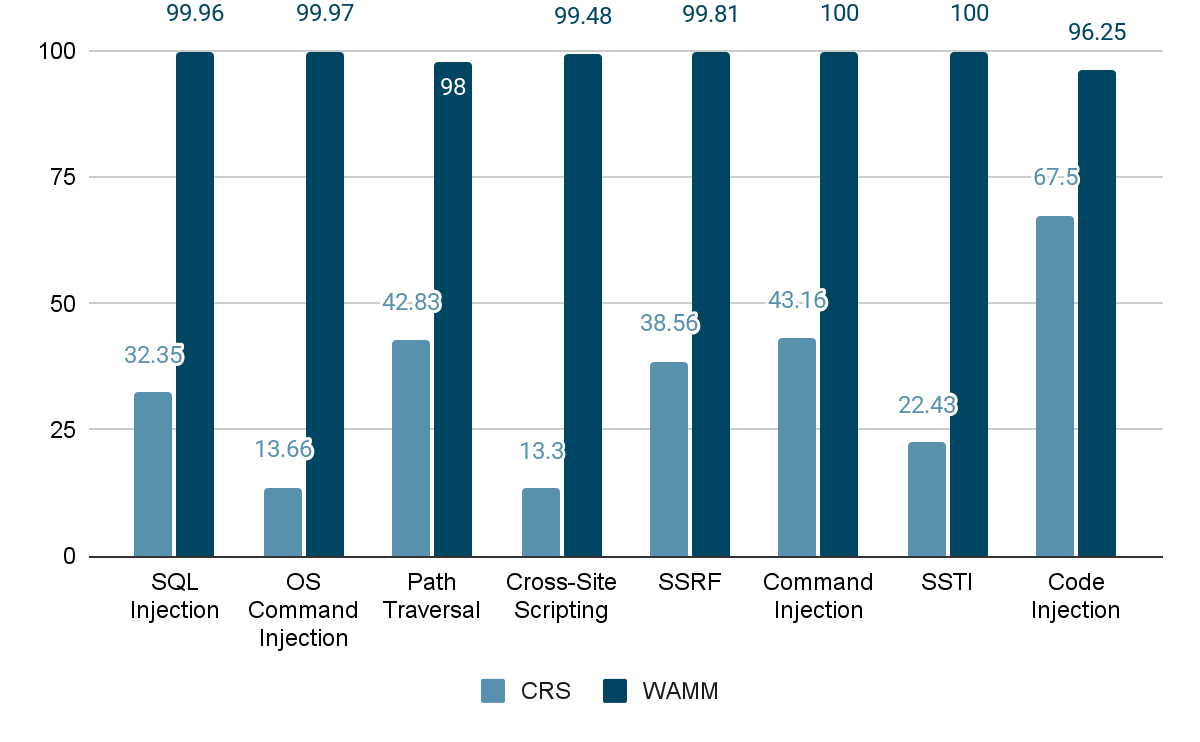}
\caption{CRS blockrate against WAMM}
\label{fig9}
\end{figure}

WAMM accuracy was further evaluated by comparing it with the OWASP Core Rule Set for the tested technology stack, which is considered the gold standard of the field of web attack classification \cite{ref62}. Both classifiers were deployed on the same test environment and assessed using the same input data, The DS-Augmented-v2-Test data served as the unseen portion of the augmented dataset. The evaluation was executed under consistent conditions, and the results are presented in Table \ref{tab7}. Which shows that across all web attack classes WAMM outperformed CRS in true positive block rates. Figure \ref{fig9} also highlights WAMM’s superior accuracy compared to CRS on the test dataset.

\begin{table*}[!t]
\centering
\caption{Summary of Detailed Classification Report}\label{tab7}
\begin{tabular}{p{2cm}lp{2cm}llp{1cm}}
\toprule
\multicolumn{2}{c}{Web Attack Class} & samples & \multicolumn{3}{c}{Block Rate} \\
ID & Name & & CRS & WAMM & $\Delta$ \\
\midrule
66 & SQL Injection & 29217 & 32.35 & 99.96 & 67.61 \\
88 & OS Command Injection & 7223 & 13.66 & 99.97 & 86.31 \\
126 & Path Traversal & 3544 & 42.83 & 98.00 & 55.17 \\
79 & Cross-Site Scripting & 579 & 13.3 & 99.48 & 86.18 \\
918 & SSRF & 529 & 38.56 & 99.81 & 61.25 \\
248 & Command Injection & 519 & 43.16 & 100.00 & 56.84 \\
1336 & SSTI & 379 & 22.43 & 100.00 & 77.57 \\
94 & Code Injection & 240 & 67.5 & 96.25 & 28.75 \\
\bottomrule
\end{tabular}%
\end{table*}

To deepen our analysis, a head-to-head comparison was conducted to evaluate WAMM against OWASP CRS using an unseen portion of the augmented dataset, using per-class true positive block rate as the endpoint. WAMM consistently outperformed CRS across all attack classes, with improvements ranging from 28.75\% (Code Injection) to 86.31\% (OS Command Injection). Significant improvements were also observed in XSS (86.18\%), SSTI (77.57\%), SQLi (67.61\%), SSRF (61.25\%), Path Traversal (55.17\%), and CMDi (56.84\%). WAMM achieved near-perfect block rates ($\ge$ 96\% - 100\%) across all classes, while CRS rates remained substantially lower. These results suggest that WAMM’s feature engineering learning approach generalizes effectively to polymorphic and obfuscated attack payloads that static rule sets miss, while maintaining broad class coverage.

\section{Conclusion and Future Work}\label{sec6}
WAMM is an extensive framework designed for the detection and classification of web attacks using HTTP payloads within a specified technology stack. It is based on the SR-BH 2020 dataset, which has been refined through deduplication, LLM-assisted relabeling, and realistic augmentation to generate three improved, richer versions (DS-Base, DS-Augmented-v1, and DS-Augmented-v2). According to the findings, carefully curated data and model selection can significantly enhance the detection process. The XGBoost achieved 99.59\% accuracy on DS-Augmented-v2 with inference times measured in tens of microseconds. Furthermore, deep learning models performed well on clean data but only reached 80\% accuracy on noisy samples and required more computation. When compared to the OWASP CRS, WAMM successfully blocked 96 to 100\% of attacks across various classes, enhancing detection rates by up to 86\% in categories like OS Command Injection. These results show the importance of combining structured dataset design with effective learning techniques for practical and real-time detection of web attacks. When trained and tested on similar technology stacks, model performance remains optimal. However, different technologies may require separate models or additional fine-tuning to achieve comparable accuracy.

As future work, an enhancement will be implemented in this research by enabling the models to support continuous learning, which will allow adaptation to changing traffic and attack patterns without the need for complete retraining. An alternative strategy involves extending the applicability of WAMM across different technology stacks to evaluate its generalization and scalability. Furthermore, the integration of ML with traditional rule-based sets rather than considering them as separate solutions. Rules can serve as priors, weak labels, or interpretable anchors, while models add flexibility and robustness against obfuscation and zero-days threats. Finally, verifying robustness and efficiency in real-world applications will need testing performance under actual deployment situations such as varying traffic loads, parser inconsistencies, and latency constraints.

\section*{Declarations}
\begin{itemize}
\item \textbf{Competing interest:} The authors declare that they have no known competing financial interests or personal relationships that could have appeared to influence the work reported in this paper.
\item \textbf{Data availability:} Data will be made available on request.
\end{itemize}

\section*{Acknowledgments}
This work was supported by the security research and development department at Cyshield, Cairo, Egypt. The authors thank Dr. Ahmed Salem (College of Computing and Information Technology, Arab Academy for Science, Technology and Maritime Transport - AASTMT, Egypt) for his constructive feedback and guidance.

\bibliographystyle{IEEEtran}
\bibliography{bibliography}

\end{document}